# Towards Media Inter-cloud Standardization – Evaluating Impact of Cloud Storage Heterogeneity

Mohammad Aazam[1, *], Marc St-Hilaire[1]
[1]Department of Systems and Computer Engineering
Carleton University,
Canada
aazam@ieee.org, marc_st_hilaire@carleton.ca

Eui-Nam Huh[*]
[*]Department of Computer Engineering
Kyung Hee University,
South Korea
johnhuh@khu.ac.kr

*Abstract*— Digital media has been increasing very rapidly, resulting in cloud computing's popularity gain. Cloud computing provides ease of management of large amount of data and resources. With a lot of devices communicating over the Internet and with the rapidly increasing user demands, solitary clouds have to communicate to other clouds to fulfill the demands and discover services elsewhere. This scenario is called inter-cloud computing or cloud federation. Inter-cloud computing still lacks standard architecture. Prior works discuss some of the architectural blue-prints, but none of them highlight the key issues involved and their impact, so that a valid and reliable architecture could be envisioned. In this paper, we discuss the importance of inter-cloud computing and present in detail its architectural components. Inter-cloud computing also involves some issues. We discuss key issues as well and present impact of storage heterogeneity. We have evaluated some of the most noteworthy cloud storage services, namely: Dropbox, Amazon CloudDrive, GoogleDrive, Microsoft OneDrive (formerly SkyDrive), Box, and SugarSync in terms of Quality of Experience (QoE), Quality of Service (QoS), and storage space efficiency. Discussion on the results shows the acceptability level of these storage services and the shortcomings in their design.

## I. INTRODUCTION

With the ever increasing Internet traffic, digital media has persuasively surpassed traditional media; as a result, vast and long-term changes are required in the way content is shared on the Internet. In 2010, the global Internet video traffic had surpassed global Peer-to-Peer (P2P) traffic [1]. Excluding the amount of video exchanged through P2P file sharing, at the time being, Internet video has gone beyond 50 percent of consumer Internet traffic. It will further grow to 62 percent by the end of 2015. Counting all forms of video, the number will be approximately 90 percent by 2015 [2]. To meet challenges and opportunities coming along with media revolution, much sophisticated services and powerful capabilities are urgently required now.

Cloud computing has recently been foreseen as a promising as well as inevitable technology. Cloud computing platform provides highly scalable, manageable and schedulable virtual servers, storage, computing power, virtual networks, and network bandwidth, according to user's requirement and affordability. It can provide solution package for the media revolution, if wisely designed and integrated with the advanced technologies on media processing, transmission, and storage. Media management is among the key aspects of cloud computing, since cloud makes it possible to store, manage, and share large amounts of digital media. Moreover, cloud computing provides ubiquitous access to the content, without the hassle of keeping large storage and computing devices. Sharing large amount of media content is another feature that cloud computing provides. Other than social media, traditional cloud computing provides additional features of collaboration and editing of content. If content is to be shared, downloading individual files one by one is not easy. Cloud computing also caters this issue, since all the content can be accessed at once by other parties, with whom the content is being shared.

With the rapid increase in digital content and user's requirements, there comes situation when a solitary cloud is no more able to fulfill the requirements. In such cases, two or more clouds have to interoperate through an intermediary cloud or gateway. This scenario is known as Inter-cloud computing or cloud federation. Inter-cloud computing involves transcoding and interoperability related issues, which also affect the overall process of multimedia content delivery [3]. Continuing our work presented in [32], the purpose of this paper is twofold. First, we discuss what are the key components involved in media inter-cloud computing. We discuss the noteworthy issues and heterogeneities involved in this regard. Second, among the heterogeneities, we focus on storage technology and present our findings to elaborate the impact of already available storage designs, so that the point of concern could be determined for future technological standardization.

In rest of the paper, section II discusses prior studies. Section III explains media cloud, while section IV presents Media Inter-Cloud architecture. Section V is on issue and heterogeneities involved. Section VI is on storage heterogeneity. We conclude our paper in section VI.

## II. RELATED WORK

Media cloud and inter-cloud computing are still in the beginning stages and there is no standard architecture available for data communication, media storage, compression, and media delivery. Already done studies mainly focus on presenting architectural blueprints for this purpose. In Intel-HP viewpoint paper [4], industrial overview of the media cloud is presented. It is stated that media cloud is the solution to suffice the dramatically increasing trends of media content and media consumption. For media content delivery, QoS is going to be the main concern. Heterogeneous QoS requirements play a very vital role in quality management and service delivery. In this regard, we present an end-to-end QoS related framework, using Flow Label of IPv6 and Multi-Protocol Label Switching (MPLS) in [5]. To reduce delay and jitter of media streaming, better QoS is required, for which Z. Wenwu et al. [6] proposes Media-Edge Cloud (MEC) architecture. The authors state in their work that an MEC is a cloudlet which locates at the edge of the cloud. MEC is composed of storage space, Central Processing Unit (CPU), and graphics processing unit (GPU) clusters. The MEC stores, processes, and transmits media content at the edge, thus incurring a shorter delay. In turn the media cloud is composed of MECs, which can be managed in a centralized or P2P way. In terms of QoS, this proposal may be efficient, but for mobile cloud computing, where resource constraint devices require real-time energy efficient service, localized solutions are required. Furthermore, they do not present about heterogeneity in QoS requirements and transcoding. Cost-effect is also not discussed.

S. Ferretti et al. [7] present an approach to use a pair of proxy, a client proxy at the user's side and a server proxy at the cloud side, to integrate the cloud seamlessly to the wireless applications. Proxies are not capable enough to handle inter-cloud in all respects. There has to be a broker which not only performs inter-cloud communication, but also, match-making and resource management. Broker also advertises the services. We present the architecture of broker in detail in this paper. D. Diaz Sanchez et al. also present proxy as a bridge, for sharing the contents of home cloud to other home clouds and to the outside public media clouds [8], [9]. Proxy can additionally index the multimedia content, allowing public cloud to build search database and content classification. Media cloud can then provide discovery service to the users to search the content of their interest. H. Zixia et al. [10] also present a proxy scheme for transcoding and delivery of media. On the other hand, J. Xin et al. [11] propose usage of P2P for delivering media stream outside the media cloud. In both the cases, it builds a hybrid architecture, which includes P2P as well as media cloud.

Transcoding and compression of media content requires a lot of resources. R. Pereira et al. present an architecture, in which Map-Reduce model is applied for this purpose, in private and public clouds [12], [13]. The authors do not consider how storage heterogeneity influence the performance.

J. Feng et al. [14] proposed the concept of stream oriented cloud and stream-oriented object. The authors introduce stream-oriented cloud with a high-level description.

Bhaskar et al. present [15] that cloud computing still faces some open issues, in terms of: scalability, availability, security, privacy, service level agreement, trust, interoperability, data migration, and resource management. Andrés García García et al. discuss in [16] about lack of proper and formal representation of storage and retrieval of cloud services. Even though they present on Service Level Agreement (SLA), but they do not present any architectural guidelines on storage and retrieval.

Kan Yang et al. present a dynamic auditing protocol for ensuring the integrity of stored data in the cloud. They present an auditing framework for cloud storage [17]. Wang Cong et al. present their work on auditing for security and privacy in cloud storage [18], [19]. Their study lack the discussion on different storage technologies and their impact on performance. Yau et al. [20] state that data integrity is among the main concerns cloud providers have. Their study focus on data integrity in clouds. Storage aspects are not discussed in their study. Zhifeng Xiao et al. [21] also discuss security and privacy concerns in cloud computing, not going into the details of overall architectural requirements as well as storage design. Talal Alsedairy et al. [22] present network densification strategy through self-organizing cloud cells. The authors state that due to massive increase in mobile data usage, cloud storage and other cloud oriented data has been increasing. This trend requires efficient resource management. But in their study, they focus only on network densification, not discussing difference in storage technologies and their impact.

III. MEDIA CLOUD

Meeting the consumption requirements of future has become a challenge now. The extraordinary growth in mobile phone usage, specially with smart phones along with 3G, 4G, Long Term Evolution (LTE), and LTE Advanced (LTE-A), with Multimedia Broadcast and Multicast Services (MBMS) networks and then the availability of more convenient access networks, like: Wi-Fi, WiBro, WiMax, Fiber to the home, and broadband networks, has hugely increased the production as well as communication of multimedia content. It is estimated that by 2015, up to 500 billion hours of content will be available for digital distribution. With social media, IPTV, Video on Demand (VoD), Voice over IP (VoIP), Time Shifted Television (TSTV), Pause Live Television (PLTV), Remote Storage Digital Video Recorder (RSDVR), Network Personal Video Recorder (nPVR), and other such services available more easily, users now demand anytime and on-the-go access to content. It has been estimated that by 2015, there will be 1 billion mobile video customers and 15 billion devices will be able to receive content over the Internet [4].

Media cloud helps fulfilling four major goals: ubiquitous access; content classification; sharing large amount of media; content discovery service. Since media content is produced to a great extent and very rapidly, it also requires efficient access, other than being ubiquitously accessible. Media cloud provides indexing and proper classification of content, which makes access of content easier as well as makes searching efficient. Media cloud also provide content discovery service, with which, content stored on other clouds can be accessed, after searching and negotiating licensing terms and conditions. This creates accessibility of huge amount of multimedia content and creates Cloud of Clouds (CoC) that can interoperate with each other, explained further later in this and next section.

Keeping in view the shortcomings and design considerations discussed, media cloud tasks are divided into layers, to make it more comprehensive that what kinds of things media cloud has to deal with and to what extent. Figure 1 presents an overall layered architecture of media cloud. At virtualization layer, cloud has to deal with computing virtualization, memory virtualization, and network virtualization. It also includes different virtualization technologies.

Storage layer deals with storage space virtualization and storage technology to be used for media content storage, like Network Attached Storage (NAS), Direct Attached Storage (DAS), Fiber Channel (FC), Fiber Channel over IP (FCIP), Internet Fiber Channel Protocol (iFCP), Content Addressed Storage (CAS) or Fixed Content Storage (FCS), and Internet Small Computer Systems Interface (iSCSI). Software Defined Storage (SDS), for example: OpenStack, ViPR, and HP StoreVirtual, also lie on this layer. SDS separates the storage capabilities and services from the storage hardware. It transforms storage into extensible, simple, open platform. Storage layer also has to deal with data security, privacy, and integrity. Other than this, data replication, de-duplication, and other added features are also part of this layer. Data replication is for the overall protection of stored contents, to make sure that if one copy of data is lost or corrupted, its replica exists. On the other hand, data de-duplication is for the user, which protects making unnecessary copies of the same content. Its purpose is to increase storage efficiency.

Next is the Access Layer, which has to deal with network access, whether the access network or wide area network. It also has to ensure secure communication of the data. Based on the access network and its capabilities, the quality of data to be sent is decided. One of the cloud storage services Dropbox provides an additional service, known as LANSync, through which files shared on the same LAN are accessed locally, instead of from the Dropbox server. This not only allows a very fast data transmission, but also does not burden the access network as well.

Middleware layer is to deal with encoding-decoding tasks and interoperability related things. As discussed, heterogeneous clients having different requirements access heterogeneous types and formats of data, as a result of which, transcoding and interoperability plays an important part in media delivery. That is why this layer is very crucial. Protocol translation is most of the times lossy, because of heterogeneous protocol structures. Similarly, tunneling bears overhead of encapsulation and decapsulation.

Application layer provides the User Interface (UI). UI can be in three forms, web interface, which is accessible through web browser. Client desktop application, running on the user's machine. Client mobile application, for mobile users. All these types of interfaces have different capabilities and features. For example, mobile client of storage service providers may not provide full editing features.

Media cloud provider, or cloud service provider has a business in all these services that it provides to its customers. Business layer deals with that part of media cloud architecture. The services will have different types for different customers and accessing devices. Quality and quantity of data is also considered when offering services. Different kinds of packages can be made available to the user. With the advent of mobile cloud computing and many other factors, cloud customers have very fluctuating behavior. Based on cloud customers' characteristics, pricing and billing is performed. It also includes incentives and penalties.

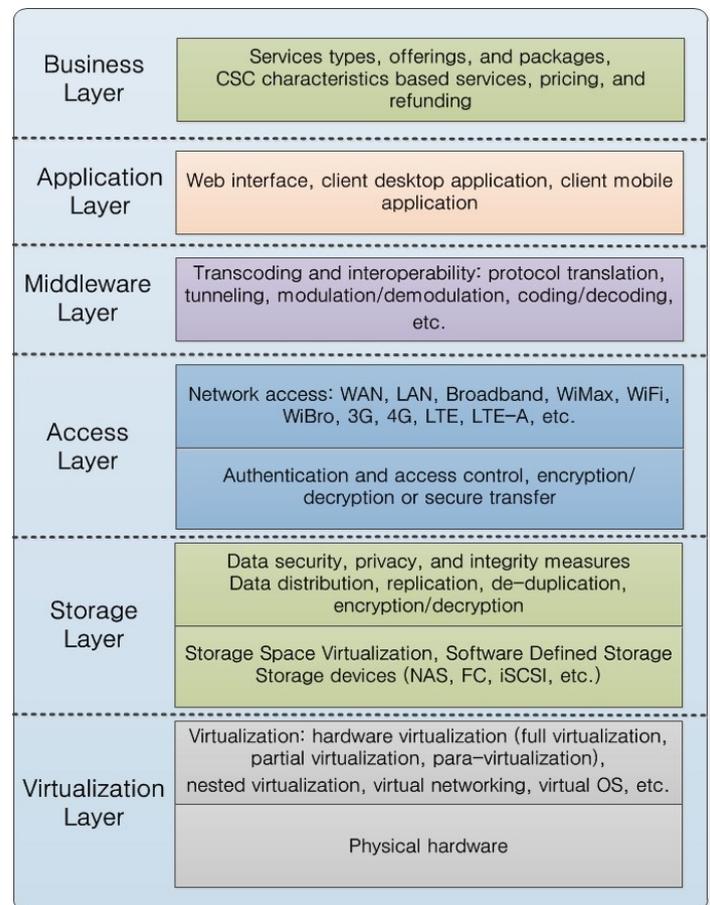

Figure 1. Media cloud layered architecture.

### A. Media Cloud design considerations

The devices receiving cloud data are of different types; hence a media cloud must have the provision to deliver content according to device's capabilities. A cloud must be able to deliver content via multiple paths, having support for multiple tenants and allowing multiple service providers to share the infrastructure and software components. When there are multiple tenants, their need keeps on changing, which media cloud should be able to meet. Cloud architecture should be able to add or remove virtual machines and servers quickly and cost-effectively. Same is the case with storage capacity. Low latency transcoding, caching, streaming, and delivery of content are must for media cloud. Disk I/O subsystem speed is going to be crucial in this regard. Using advanced technologies, like Solid State Drives (SSD), Serial Attached SCSI (SAS) interfaces, next-generation processors, etc., would become a necessity.

Power utilization is another vital aspect to be considered. Due to huge amount of processing and communication of large amount of data, which is then received by devices, including small, power constraint nodes, it is going to be very important to have an affective power saving mechanism. Since we are talking about media cloud, which involves Virtual Machines (VMs), which are created during run-time to suffice user's needs. At times, many VMs will not be in use, whether temporarily or permanently, as a result, they should be monitored and suspended or shutdown, according to situation. This will save a lot of power of the datacenter. On the receiver's side, data received should only be what is required and should be in the most appropriate format. It should be according to the receiving node's requirements as well as processing and power utilization attributes. For thin client, the device which has to perform all these activities, like, broker or access network gateway, must ensure presentation conformity of the client node. The forecast, based on the power consumption trend says that by year 2021, the world population would require 1175GW power to support media consumption [4].

Security and protection is going to be another issue [18], [19], [23]. Data, receiver, and VM security would become difficult to manage, though very important. In a virtual networking and multitenant environment, VM isolation and isolation of clients becomes very important. Similarly for data, it is not only required to store the data protectively and securely, but also to be transmitted through secure session. In other words, both kinds of security: storage security and communication security are required.

Current state-of-the-art devices can produce, store, and deliver high quality media content, that can be further shared on social media and other media forums. Since different types of digital media contents can be produced and disseminated across different networks, therefore a standard mechanism is required to allow interoperability between clouds and transcoding of media contents [9]. Purpose of media cloud is to address this problem and to allow users constitute a cloud and manage media content transparently, even if it is located outside the user's domain. Different device types, resolutions, and qualities require generating different versions of the same content. This makes transcoding one of the most critical tasks to be conducted, when media traverses networks. This requires a lot of media processing that is computationally expensive.

IV. MEDIA INTER-CLOUD ARCHITECTURE

Service providers have their customers dispersed all around the globe. To serve them optimally, service providers have to setup many of their datacenters at different geographical locations. Existing systems are not capable enough to coordinate dynamically the load distribution among data centers, to determine optimal location for hosting services to achieve desired performance. Moreover, users' geographical distribution cannot be predicted as well. Hence, load coordination and service distribution has to be done automatically. Inter-cloud computing is meant to counter this problem. It provides scalable provisioning of services with consistent performance, under variable workload and dynamically changing requirements. It supports dynamic expansion and contraction of resources, to handle abrupt variations in service demands [24]. High availability and cost reduction are also among the factors for which inter-cloud computing is required [25]. Many workflow applications require specific infrastructure requirements. It is not always possible for a single cloud to fulfill the required computing and storage resources demand. Cloud customers are looking for better service at an affordable price. All such scenarios motivate the advent of inter-cloud computing [26].

With inter-cloud computing, where multiple independent cloud service providers cooperate with each other, a customer's request from one service provider is entertained by another service provider, through the mediation of cloud broker [27]. Broker's responsibility is to identify appropriate service provider, according to the needs of its customer, through cloud exchange. Broker negotiates with the gateway to allocate resources, according to user and service requirements [24]. For this purpose, cloud interoperability must be in a standardized way. Standardized way of SLA must be made part of it. Inter-cloud Protocol, with the support of 1-to-1, 1-to-many, and many-to-many cloud to cloud communication and messaging must exist [28].

*A. Inter-cloud Entities*

To start with, first the entities are to be defined. Four entities are involved in Inter-cloud communication, as explained below:

*1. Cloud service provider*

Cloud Service Provider (CSP) provides cloud services to the Cloud Service Customer (CSC), Cloud Service Partner, and other Cloud Service Providers. Provider may be operating from within the data center, outside, or both. Cloud Service Provider has the roles of: cloud service administrator, cloud service manager, business manager, and security & risk manager.

Cloud service administrator has the responsibility of performing all operational processes and procedures of the cloud service provider. It makes it sure that services and associated infrastructure meets the operational targets.

Cloud service manager ensures that services are available to the customers for usage. It also ensures that services function correctly and adhere to the SLA. It also makes sure that provider's business support system and operational support system work smoothly.

Business manager is responsible for business related matters of the services being offered. Creating and then keeping track of business plans, making service offering strategies, and maintaining relationship with the customers are also among the jobs business manager performs.

Security & risk manager makes it sure that the provider manages the risks appropriately, which are associated with deployment, delivery, and usage of the offered services. It includes ensuring the adherence of security policies to the SLA.

The sub-roles of CSP include: inter-cloud provider, deployment manager, and customer support & care representative.

Inter-cloud provider relies on more than one CSPs to provide services to the customers. Inter-cloud provider allows customers to access data residing in external CSP by aggregating, federating, and intermediating services of multiple CSPs. It adds a layer of technical functionality that provides consistent interface and addresses interoperability issues. Inter-cloud provider can be combined with business services or independent of this.

Deployment manager performs deployment of service into production. It defines operational environment for the services, initial steps, and requirements for the deployment and proper working of the services. It also gathers the metrics and ensures that services meet SLA.

Customer support & care representative is the main interface between customers and provider. Its purpose is to address customer's queries and issues. Customer support & care representative monitors customer's requests and performs the required initial problem analysis.

### 2. Cloud service customer

CSC is that entity which uses cloud services and has a business relationship with the CSP. The roles of CSC are as: cloud service user, customer cloud service administrator, customer business manager, and customer cloud service integrator.

Cloud service user only uses cloud service(s), according to the needs. Customer cloud service administrator ensures that the usage of cloud services goes smoothly. It oversees the administrative tasks and operational processes, related to the use of services and communication between the customer and the provider.

Cloud business manager has a role of meeting business goals of customer, by using cloud services in a cost effective way. It takes into account the financial and legal aspects of the usage of services, including accountability, approval, and ownership. It creates a business plan and then keeps track of it. It then selects service(s), according to the plan and then purchases it. It also requests audit reports from the Auditor, an independent third party, explained further below, in subsection 3.

Customer cloud service integrator integrates the cloud services with customers internal, non-cloud based services. For smooth operations and efficient working, Integrator has a very vital role to play. Services' interoperability and compatibility are the main concerns in this task.

### 3. Cloud service partner

Cloud Service Partner is kind of a third party which provides auxiliary tasks, which are beyond the scope of CSP and CSC. Cloud Service Partner has the roles of Cloud Developer, Auditor, and Cloud Broker.

In a broad sense, Cloud Developer develops services for other entities, like CSC and CSP. Among the main roles, Cloud Developer performs the tasks of designing, developing, testing, and maintaining the cloud service. Among the sub-roles, Cloud Developer performs as Service Integrator and Service Component Developer.

As Service Integrator, Cloud Developer deals with composition of service from other services. While as a Service Component Developer, it deals with design, development, testing, and maintenance of individual components of service. Cloud Developer ensures meeting the standard of development, based on certain user or general users, according to the requirements of the project. Since Inter-cloud computing is going to be standardized, it should be mentioned in case of Cloud Service Developer that the services being developed must meet the standard. As heterogeneous clients (devices) are going to use the services and on the other hand, various diverse types of development environments are available to the developer, hence, it must be tightly coupled with some specific standard of development.

Cloud Auditor performs the audit of the provision and use of cloud services. Since service provider and service customer are separate entities, therefore, the service quality, usage behavior, and conformance to SLA, all this has to be audited by the third party, having the role of Auditor. Audit covers operations, performance, security, and examines whether a specific criteria of SLA and the audit is met or not. Auditor can be a software system or an organization.

Cloud Broker is the intermediary, through which two or more clouds are interoperated and their resources are federated. It is also referred to as Inter-cloud broker, or simply, broker.

Cloud broker provides a single interface through which multiple clouds can be managed and share resources [29]. Broker operates outside of the clouds and controls and monitors those clouds. The main purpose of the broker is assisting the customer to find the best provider and the service, according to its requirements, with respect to specified SLA. Broker provides its customers with a uniform interface to manage and observe the deployed services. Broker earns its profit by fulfilling requirements of both the parties. It uses a variety of methods, such as a repository for data sharing and integration across data sharing services to develop a commendable service environment and achieve the best possible deal and SLA between two parties, i.e., CSP and CSC [30] [31]. Broker typically makes profit either by taking remuneration from the completed deal or by varying the broker's spread, or some combination of both [32]. The spread is the difference between the price at which a broker buys from seller (provider) and the price at which it sells to the buyer (customer).

To handle commercial services, broker has a cost management system. As shown in Figure 2, cloud broker includes Application Programming Interfaces (APIs) and a standard abstract API, which is used to manage cloud resources from different cloud providers. Broker holds another abstract API for the negotiation of cloud service facilities with the customer. Different modules perform a specific task in broker's architecture, e.g., registration of new services is handled by Service Registration Manager. Deployment of services and making them available is done by Deployment Manager. Similarly, each module has its own specific utility.

CSC can directly access CSP(s) as well. But in that case, transcoding related tasks, SLA negotiation, and match-making is done by CSC itself.

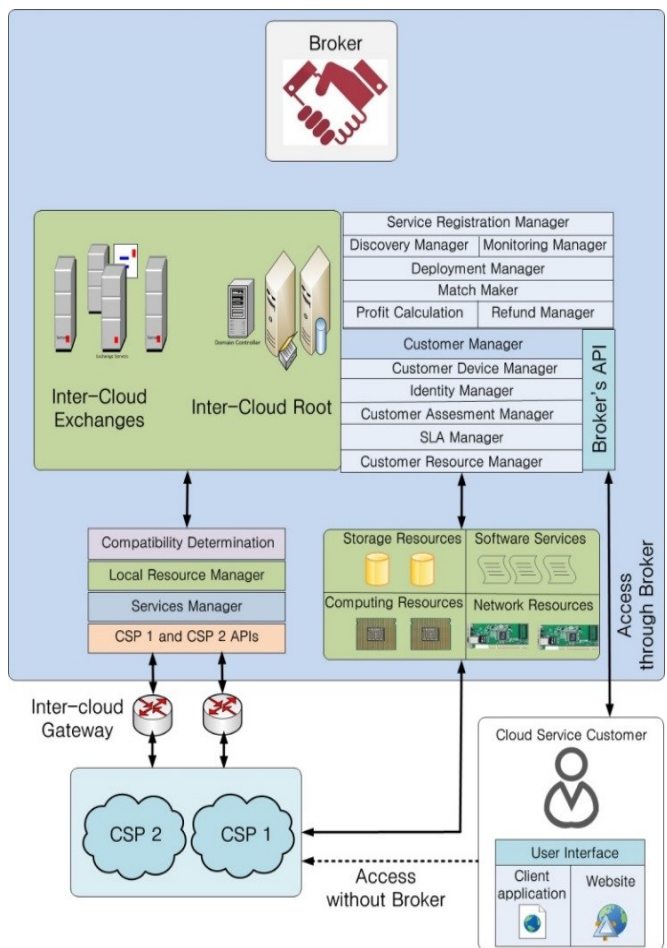

Figure 2. Broker's architecture and communication scenarios.

*4). Cloud service carrier*

Cloud service carrier is an intermediary that provides connectivity and transport of cloud services, from CSP to CSC. With the role of Cloud Network Provider it provides network connectivity and related services. It may operate within the data center, outside of it, or both. It provides network connectivity, provides other network related services, and manages the services.

Figure 3 presents the entities and their roles in Inter-cloud computing or cloud federation.

### B. Inter-cloud Topology Elements

Inter-cloud computing involves three basic topology elements, which are explained in this part.

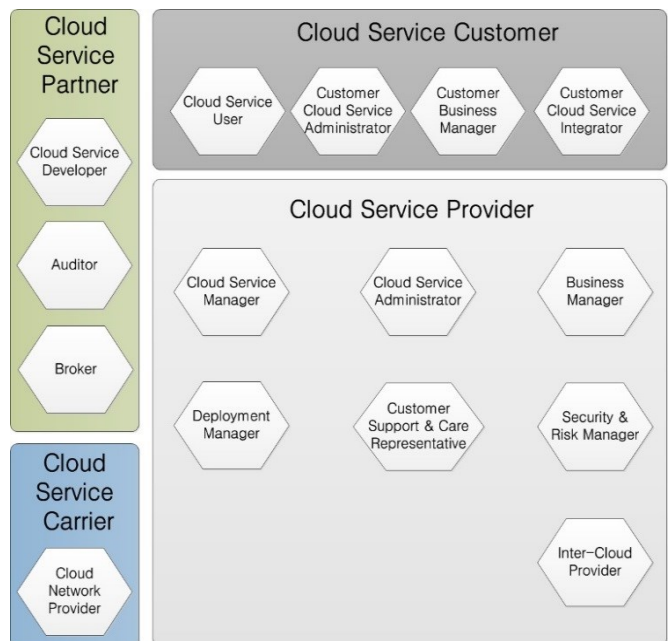

Figure 3: Inter-cloud computing entities, their roles, and sub-roles.

#### 1. Inter-cloud Exchanges

Inter-Cloud Exchanges (ICXs) are those entities which are capable of introducing attributes of cloud environment for inter-cloud computing. It is a complex and variable system of service providers and the users of services. ICX is a mediator, which brings together service providers and customers. Its responsibility is to aggregate infrastructure demands from the broker and match them against the resources available, published by the Gateway. It involves applications, platform, and services needed to be accessible through uniform interfaces. Brokering tools play an important part in actively balancing the demands and offerings, to guarantee the required SLA at higher levels of service. CSPs exchange resources among each other effectively pooling together part of their infrastructure. ICXs perform aggregation of offer and demand of computing resources, creating an opportunity for brokering services. In ICXs, proxy mechanisms are required to handle active sessions, when migration is to be performed. This is done by Redirecting Proxy in Inter-Cloud Exchange. Redirecting Proxy performs public IP to private IP mapping. This public IP to private IP mapping is important to provide transparent addressing.

#### 2. Inter-cloud Root

Inter-cloud Root contains services like, Naming Authority, Directory Services, Trust Authority, etc. Naming Authority is responsible for nomenclature related matters. Naming is an important matter in cloud storage, since every single file has a unique web-ID. It basically provides two different functions. First, it provides the scope of the name, making sure that the name is unique within its current scope. Second, it provides the format of the name by breaking the name into name fields, through delimiters. Directory Service, as the name suggests, provides mean for storing, organizing, and providing access to the information. It identifies all the resources in the datacenter and makes them accessible. Trust Authority deals with how transactions and communication take place securely. Digital certificates, security parameters, model of trust relationship, cloud infrastructure security, etc. are all part of it. Inter-Cloud Root is physically not a single entity, but a DNS-like global replicating and hierarchical system. It may also act as broker.

### 3. Inter-cloud Gateway

It is a router that implements Inter-cloud protocols and allows Inter-cloud interoperability. It provides mechanism for supporting the entire profile of Inter-cloud protocols and standards. Once the initial negotiation is done, each cloud collaborates with others directly. The purpose of Inter-Cloud Gateway is providing mechanism for supporting the entire profile of inter-cloud standards and protocols. On the other hand, the Inter-Cloud Root and Inter-Cloud Exchanges mediate and facilitate the initial inter-cloud negotiation process among Clouds. Cloud Gateway is a coordinator between internal datacenters and external clouds. Its responsibility is to publish the services to the cloud federation.

Figure 4 presents Inter-cloud elements and Inter-cloud Provider architecture.

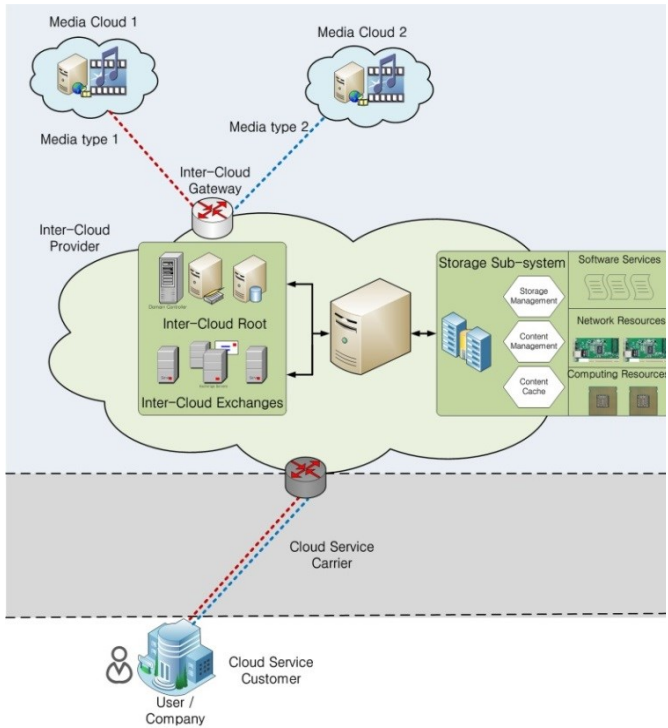

Figure 4. Inter-cloud scenario and topology elements.

## V. ISSUES AND HETEROGENEITIES

Being still in its infancy, media cloud faces a lot of challenges and heterogeneities, which have to be addressed first, before it is formally standardized. There are several issues and heterogeneities involved with inter-cloud computing. Our purpose here is to briefly highlight the noteworthy concerns here. We elaborate in detail on one of the key concerns in the next section.

### 1) Heterogeneous media contents and media transcoding

Very diverse types of services are available in the media cloud arena, making transcoding and content presentation an area of concern. Services like, VoD, IPTV, Voice over IP VoIP, TSTV, PLTV, RSDVR, nPVR, and the increasing social media content requires a lot of effort in this regard.

### 2) Heterogeneous access networks

Every access network, like, broadband, WiFi, WiBro, GPRS, 2G, 3G, 4G, LTE, LTE-A, and other upcoming standards have different attributes, bandwidth, jitter tolerance, and performance. Service quality is directly influenced by the type and capabilities of access network, other than the condition of core network. User's mobility is also among the factors. Cloud datacenters have to manage resources on the basis of type of access network.

### 3) Heterogeneous client devices

When contents are available on cloud, any device that has access to the Internet can request for service. All types of client nodes have different capabilities and constraints. Other than the type of contents it can support, the size of display, buffer memory, power consumption, processing speed, and other such attributes have to be considered before fulfilling the request. With the arrival of Internet of Things (IoT), different types of sensors and devices, and nodes are able to access cloud services. Many of the IoT based services are envisioned today to be based on cloud computing, creating Cloud of Things (CoT). For all such cases, accessing nodes, their types, capabilities, type of data they require, type of data they generate (specially in the case of CoT) plays a very vital role in media and resource management as well as service delivery We discuss it in detail in [33].

### 4) Heterogeneous applications

Requesting applications are also of different types and require different treatment. Other than the heterogeneity of device, the application type also matters. For example, web browser requesting service will have different requirements, while cloud client application will have different requirements for the same service. Some of the mobile clients do not provide all those features which desktop client provides. Cloud storage services also lack some editing features in mobile cloud client.

### 5) Heterogeneous QoS requirements and QoS provisioning mechanisms

Depending upon the access network, condition of core network, the requesting device, user's needs, and type of service, heterogeneous QoS requirements can be made. Dynamic QoS provisioning schemes are required to be implemented in this regard. We have worked on it in detail and present end to end QoS provisioning mechanism, using flow label of IPv6 and Multi-Protocol Label Switching (MPLS) [5].

### 6) Data/media sanitization

When a client requests for storage space from the cloud, it does not mean that 'any' type of data can be now stored. Data has to be filtered. Filtration may be based on moral grounds, some specific laws, or political grounds. Some of the cloud storage service providers do not allow some specific type of data to be stored, like pornographic material. One of such services is Microsoft OneDrive (SkyDrive), which scans the data for storage, for this particular purpose.

*7) Security and trust model*

Outsourced data poses new security risks in terms of correctness and privacy of the data in cloud. When we talk about media cloud, not only data service will be requested by the user, but also, storage service would be requested too. Storing contents, which may have some sensitive or private information, poses risks to the customers. When data is stored in an outsourced resource, it is no more under the protection of its owner. Not only data integrity becomes a concern, but privacy as well becomes an issue. On Feb 01, 2013, it was read on The Independent, stating, "British internet users' personal information on major 'cloud' storage services can be spied upon routinely by US authorities"[1]. We provide further details in [34].

*8) Heterogeneous Internet Protocols*

IPv4 address space has exhausted. Migration towards IPv6 has formally been expedited. Both of these versions of IP are not directly interoperable. Since this migration is going to take some time, may be a decade, meanwhile, they have to be made interoperable. Tunneling is the viable solution in hand for this purpose, but it has its own overhead. We have worked extensively on this and presented our findings in [35].

*9) Heterogeneous media storage technologies*

Storage is an important part. Multimedia content require a lot of space. Moreover, in case of multimedia contents, it becomes more difficult to search on the basis of actual contents the file contains. Efficiency in storage and searching is an important aspect media cloud should have. Different storage technologies available are: NAS, DAS, FC, FCIP, iFCP, CAS or FCS, and iSCSI. Communication between clouds create inefficiency when different storage technologies are provided by the service providers. When transcoding is performed, it is primarily lossy, which degrades the quality. Some of the existing work uses simple storage schemes for multimedia content, while most of the works rely on Hadoop Distributed File System (HDFS) [36]. But the issue with HDFS is that it is designed mainly for batch processing, rather than for interactive user activities. Likewise, HDFS files are write-once and can have only one writer at a time, which makes it very restricted for those applications which require real-time processing, before the actual delivery of data. Other than these issues, dynamic load-balancing cannot be done with HDFS, since it does not support data rebalancing schemes. In section VI, we have discussed about the impact of different storage techniques, adopted by cloud storage services.

---

[1] http://www.independent.co.uk/life-style/gadgets-and-tech/news/british-internet-users-personal-information-on-major-cloud-storage-services-can-be-spied-upon-routinely-by-us-authorities-8471819.html.

Other than the existing heterogeneities and the ones that are to be emerged, media cloud needs to be able to deal with dramatically increasing video contents. Until 2015, in every second, 1 million minutes of video content will cross the network [2]. Therefore, it is very important to carefully design the architecture of media cloud, to be a successful platform and to be able to adapt to continuously increasing amount of media content, new applications, and services.

## VI. STORAGE HETEROGENEITY

In this section, we discuss the impact of heterogeneity in storage technology, by presenting performance evaluation of noteworthy cloud storage services, on certain parameters.

### A. Evaluation and Setup

There are different types of storage technologies that are being used by different cloud storage services. Storage technology has to be standardized to ensure efficiency of coding-decoding and storage space. In case of proprietary systems, internal details remain unknown. But when we evaluate, we can at least get performance measures on some particular parameters and conclude that on what grounds a particular service is performing well or underperforming. In this section, we present the impact on storage and quality of data, because of heterogeneous cloud storage technologies.

For this evaluation, we used six noteworthy cloud storage services, namely: Dropbox, GoogleDrive, Amazon CloudDrive, Microsoft OneDrive (SkyDrive), Box, and SugarSync. In the first part, the evaluation is done in Korea. Since CSC and CSP datacenter's location also matters, therefore, we are also going to present results from other locations as well in future.

The setup is in this way that a test computer is located in Korea. From different locations (Campus 1, Campus 2, Home 1, and Home 2), the statistics were gathered and their results were averaged. In all cases, same device was used. Test computer is responsible to create files, upload them to each of the cloud storage service provider.

*Data set:* The data set has two types, (i) 15 MB HD video and (ii) 10MB bulk data. The purpose of using HD video is to let cloud services use their maximum resources and to allow user analyze QoE in a better way. If the video is non-HD, it would be very difficult to judge the difference in quality of each service. Bulk data set enabled us to analyze how different types, formats, and sizes of files are handled by the cloud storage service. Different scheduling algorithms, like First-In-First-Out (FIFO), Shortest-Job-First, etc., put different impact on data upload and synchronization efficiency.

*Parameters:* Parameters were selected keeping in view the extent interaction of CSCs with cloud storage services. Which includes: upload delay, download delay, synchronization delay, jitter, bulk-data upload delay, bulk-data download delay, bulk-data synchronization delay, storage space efficiency, and some other parameters explained within these parameters in this section later.

Figure 5 shows the way setup was created.

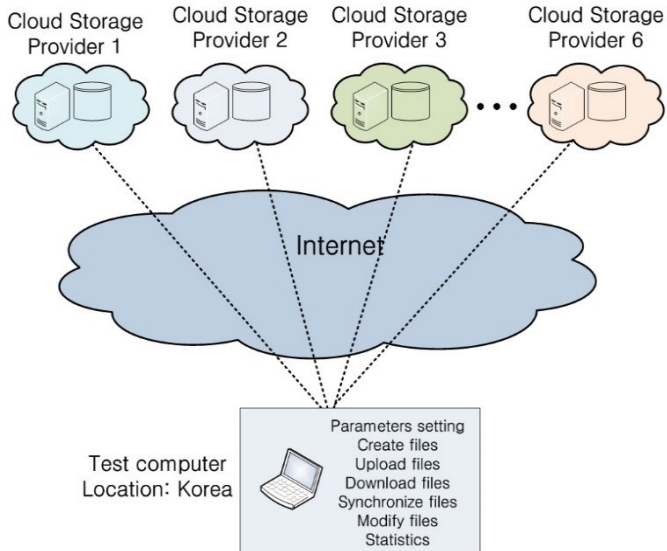

Figure 5. Upload Delay comparison of Cloud Storage Services.

A lot depends upon the network condition and the current load on storage server, when the results are being gathered. At times, servers are too busy due to user's requests and handling large amount of data, while on some other occasion; server might be quick in response. Besides, it is possible that user's trend may be different on weekends and weekdays. To ensure the reliability of results, we gathered multiple samples during different times of the day, on weekdays as well as weekends. This process of gathering results was stretched to around six weekends and up to six weeks. In this way, network conditions and current server load is normalized and results gathered are more realistic.

*1) Upload delay*

As discussed, different scheduling algorithms used by the cloud storage services have different impact on data manipulation. There is a difference in performance of each cloud storage service in uploading exactly the same content. To make sure that temporary condition of server and network condition does not affect the results, we conducted this evaluation several times, together for each service. The averaged results are presented in Figure 6. It took SugarSync around 16 seconds to upload 15MB data. It shows that SugarSync has the most efficient technique to upload contents quickly. On the contrary, GoogleDrive and Amazon CloudDrive remained slow in this regard, taking 80 and 78 seconds respectively. We did this evaluation on larger sized data sets as well, up to 1GB, but delay ratio remained the same, that is why those results are not presented here, as they do not give any additional information.

*Data Compression, prior to upload:*

For Dropbox and GoogleDrive, the reason for this delay is that these services compress the data before uploading it. It has both positive and negative aspects. For larger sized data, this compression sheds a reasonable amount of load from the core network. On the other hand, CSC has to wait longer for files to get uploaded in the cloud. Also, for images, CSCs does not want them to be compression, since it degrades the quality. Furthermore, CSC's resources are also consumed for this purpose, when data is compressed prior to upload. CloudDrive does not provide any compression before uploading the files, but its client's weaker capabilities are the reason behind the delay.

*Chunking:*

Chunking refers to splitting files into pieces and then uploading them. For large sized files, chunking plays an important role in efficiently uploading the content. Moreover, when files are not uploaded in a single attempt, later on, remaining part can be uploaded, if the service is equipped with chunking feature. If chunking is not provided, then on every attempt, the cloud storage service has to start altogether a new session for uploading the files from the beginning. While gathering the statistics, monitoring the pause in upload implies that the service performs chunking. One of the reasons of CloudDrive's slow behavior is that it does not support chunking. All other services perform chunking in a variable way. Still, GoogleDrive has the worst performance in this regard. The reasons, as stated above, is higher overhead of compression. Among all the services, GoogleDrive has the smartest compression mechanism, which makes it slow in uploading the data.

*Deduplication:*

For the sake of storage efficiency, cloud storage service providers have to prevent de-duplication, in which, same file is prevented from uploading again. In case CSC makes copy of an already uploaded file, only meta-data is updated. In this regard, only Dropbox provide this facility.

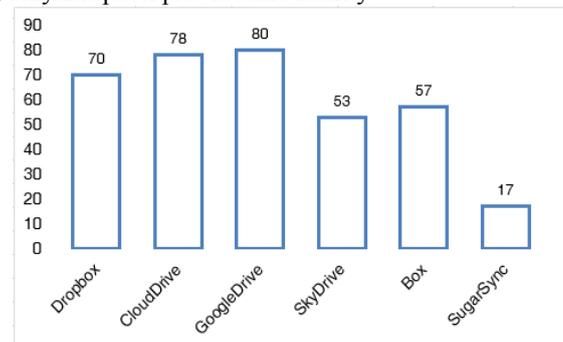

Figure 6. Upload Delay comparison of Cloud Storage Services.

*2) Download delay*

Download efficiency is determined by the server's capability to search, retrieve, and process the request. Based on the underlying storage technology and searching techniques, download is affected. Also, data integrity measures, protection (encryption/decryption), compression/decompression, etc. also play their part in this regard. In Figure 7, Microsoft's OneDrive (SkyDrive) has the worst performance in this regard, taking 50 seconds to deliver a 15MB file. This is primarily because of control signaling. GoogleDrive and SugarSync remained most efficient in this regard, with 10 and 12 seconds delay respectively. CloudDrive being morel lightweight service, only provide synchronization facility, i.e. no download is possible. Therefore, it could not be evaluated in this parameter.

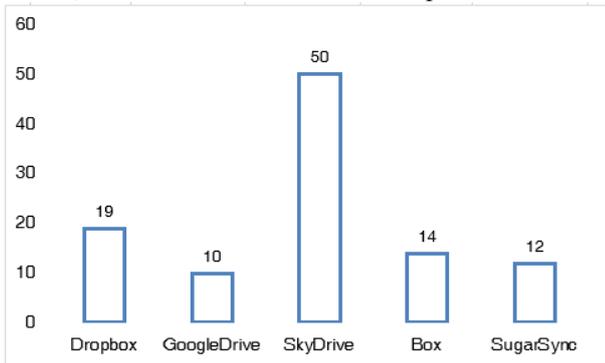
Figure 7. Download Delay comparison of Cloud Storage Services.

*3) Synchronization delay*

Synchronization delay plays an important role in the overall efficiency of a cloud storage service, when files are moved within the cloud or when collaboration work takes place between two or more parties. Any file in the cloud that can be shared has a unique web-address. When one or more files are moved from one location to another, their URLs are re-generated and updated in the cloud server. New location is also synchronized with other devices linked to the user's account. It tells that how quickly a service can index files and if some file is renamed or its location is changed, how much time it takes to the server to synchronize it. Moreover, in case of collaborative work, synchronization delay plays an important role, as multiple users are working on the same document and quick update of their activities matter a lot. Figure 8 shows Synchronization Delay.

Box had the least acceptable performance. On the other hand, Amazon CloudDrive, SugarSync, and Dropbox came up to be the most efficient ones. It was observed during this study that when a file is moved from one folder to another, Box uploads the whole file again to the new location, causing more delay. Other cloud storage services presented here does not do it this way. They just update the new location, instead of moving the whole contents physically.

*Delta Encoding:*

In the case when file is not only relocated or renamed, but also, it is modified, then new content has to be synchronized. In this regard, Dropbox is the most efficient one, since it uploads only the updated part. All others upload the whole file again, which incurs a lot of delay and overhead. In case of collaborative work, this feature affects the performance a lot.

*Polling:*

Cloud storage service has to periodically check for any update in the existing files, so that the renewed content could be updated in the cloud as soon as possible. It has to be handled with a trade-off. Quick polling not only consumes resources and network bandwidth, but also keeps storage client busy. But it senses the change soon. On the other hand, slower polling intervals does not burden the storage client and saves resources. In this regard, Dropbox's polling interval is 5 minutes. Previously it was 1 minute. In case of OneDrive (SkyDrive), it has to contact many Microsoft Live servicer, which creates an overhead. CloudDrive has the shortest polling interval, an average of about 15 seconds. This means that after every 15 seconds, polling packets are exchanged. This affects to a reasonable extent when it comes to cellular network packages (3G, 4G).

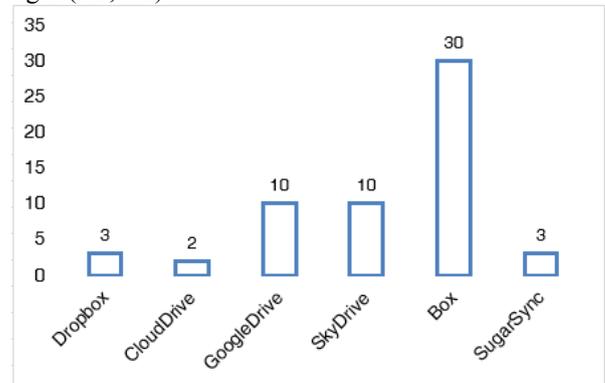
Figure 8. Synchronization Delay comparison of Cloud Storage Services.

*4) Jitter*

Figure 9 depicts the QoS and QoE aspects of the cloud storage services we used in our study. It shows that how each cloud storage service performs in terms of Jitter. Box shows the least amount of Jitter, when HD video was played-back. On the contrary, Dropbox had the highest Jitter. Amazon CloudDrive does not allow file download, as a result, its jitter could not be measured. In future, multimedia content would be much more than other types of data. Such QoE and QoS aspects are going to be even more important.

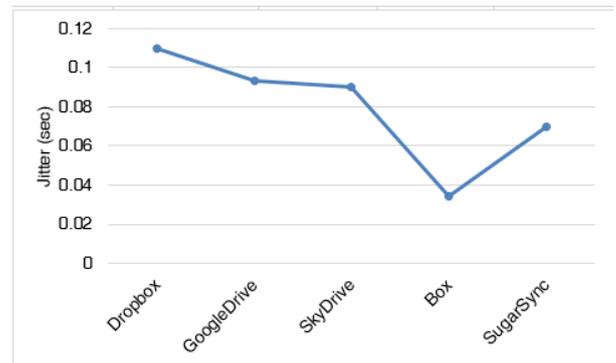
Figure 9. Jitter comparison of Cloud Storage Services.

*5) Bulk-data upload delay*

Cloud storage services have different mechanisms to handle different types of files, having different sizes and formats. Number of TCP connections established according to file upload also matters here. Delay shown in Figure 10, GoogleDrive, SugarSync and CloudDrive open TCP connections individually for each file that is to be uploaded. SkyDrive wait for the application layer acknowledgment after uploading each file.

*Bundling:*

When multiple files are to be uploaded, combining them and uploading together creates efficiency in total upload delay. Services that do not implement file bundling strategy have more overhead in uploading bulk-data files.

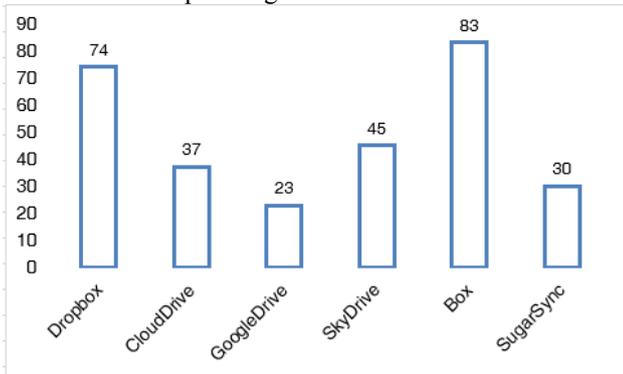

Figure 10. Bulk-data Upload Delay comparison of Cloud Storage Services.

*6) Bulk-data download delay*

Same as uploading bulk-data, downloading also involves mechanisms to deal with multiple heterogeneous files. Different files are stored in a different way. Hence, encoding decoding also comes into play. CloudDrive and GoogleDrive does not provide download of multiple files together, due to which, their results are not in Figure 11. Rest of the services have nothing significantly different in handling bulk-data, in terms of downloading. It takes 12 seconds to download a 10MB bulk-data of heterogeneous files from Dropbox. In case of OneDrive (SkyDrive), this delay is 15 seconds. Box and SugarSync incur 10 and 14 seconds delay, respectively.

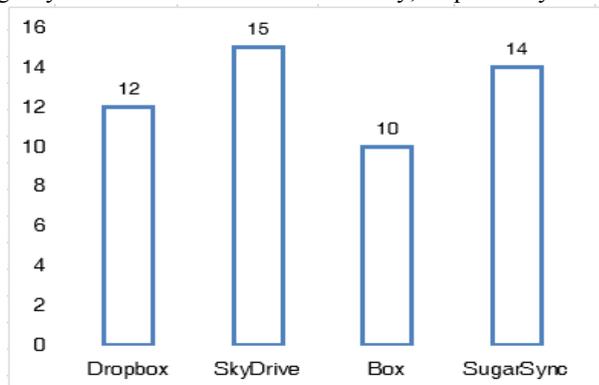

Figure 11. Bulk-data Download Delay comparison of Cloud Storage Services.

*7) Bulk-data synchronization delay*

Same as discussed in single file synchronization delay, bulk-data synchronization also has its importance, since in this case, there are multiple files. Figure 12 shows that Box has the worst performance in this parameters. Except some control signaling which has to be done when a file is uploaded for the first time, rest of the contents are entirely replaced, which is the reasons behind its comparatively worst performance. CloudDrive is most efficient because it does not provide content sharing facility and it has to only synchronize the meta-data. In other words, at the cost of sharing the content, synchronization is performed efficiently.

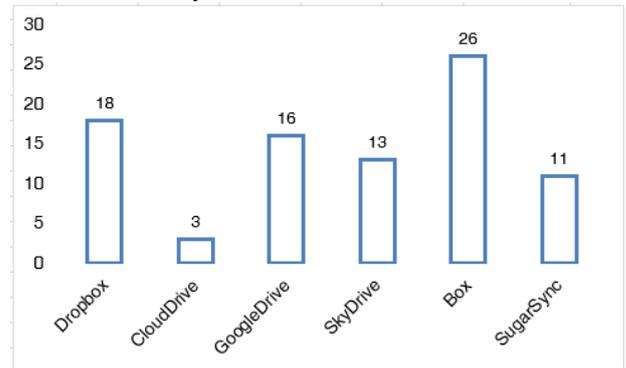

Figure 12. Bulk-data Synchronization Delay comparison of Cloud Storage Services.

*8) Storage compression and space efficiency*

With increasing cloud storage contents, efficiency in storage becomes very important. Efficient storage techniques allow cloud storage services consume minimum possible space to store data. It also makes indexing and searching efficient. In the end, it has an impact on synchronization of data and the overall efficiency of content manipulation as well. The difference created in size of storage of the same data, when stored on each cloud storage service is presented here. Figure 13 presents the size comparison on 50MB data set. It shows that Box creates 8.7% decrease in the size (compression) of stored data, which is the highest, as compared to other storage services selected in this study.

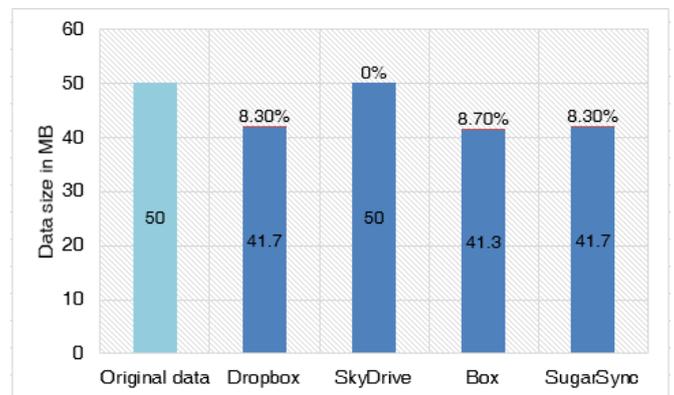

Figure 13. Storage Space Comparison on 50MB dataset.

Microsoft OneDrive (SkyDrive) does not compress the size by applying any sort of compression or storage efficiency technique. Its percentage in this regard was thence 0%. Amazon CloudDrive does not provide file download, while GoogleDrive does not allow folder download. As a result, their evaluation on this parameter could not be done. Figure 14 shows the results on 100MB data set. To ensure the reliability of results, same type and format of files were used. Only the total size of data set was increased, so that the type of file does not affect the storage size here.

In this case, although the data set has been doubled to its previous size, but the difference in storage size is not accordingly, in case of each cloud storage service. Box shows 16.3% difference in size, by decreasing the storage space consumed by the data stored. This shows an overview of how large sized data are going to affect the storage space. When the data size is increased more, storage space efficiency would not increase with that ratio. Dropbox, Box, and SugarSync use compression technique very similar to that of WinRAR.

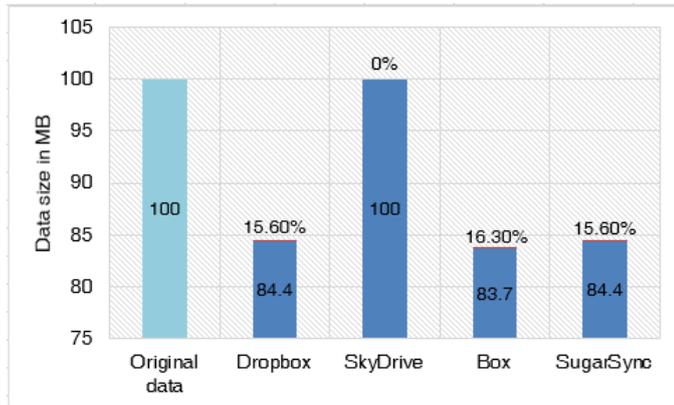

Figure 14. Storage Space Comparison on 100MB dataset.

## VI. CONCLUSION AND FUTURE WORK

In this paper, we presented the motivation for having a standardized inter-cloud computing architecture, specially in respect of media content. We provided the basics presented in the already done works and extend them as well as include additional components, according to our own research findings to present an overall picture, how media cloud should be having and what are its design considerations. We further present the noteworthy challenges media inter-cloud computing faces, so that it architecture could be standardized keeping those facts in front. In the later part, we extended the explanation of one of those issues, which was about storage heterogeneity. We present the methodology to check capabilities and system design of discussed cloud storage services. We presented the effect of this heterogeneity by evaluating performance of 6 well-known cloud storage services. This helped us elaborate the reasons behind difference in performance, caused by the underlying technologies. In this regard, it can be concluded that generally, CloudDrive comes up being the most light-weight, having least number of features. On the other hand, Dropbox and SugarSync are among the most sophisticated ones. GoogleDrive enjoys the benefits of Google's infrastructure, however, in some regards, specially prior compression, it lags behind. Box only performs well for multimedia playback. Polling, bundling, chunking, de-duplication, compression, and delta-encoding play a very vital role in the overall efficiency as well as storage of content.

To quantify long-term effects of the performance, we intend to extend our work in future by taking measurements from other locations. We would be able to present results from locations: Canada, USA, France, UK, Germany, and Singapore. We are in collaboration with Concordia University, Canada, UPMC France, Bremen University, Germany, and Bournemouth University, UK.

ACKNOWLEDGMENT

This research was supported by Basic Science Research Program through the National Research Foundation of Korea (NRF) funded by the Ministry of Education (No.NRF-2013R1A1A2013620) and the MSIP (Ministry of Science, ICT & Future Planning), Korea, under the ITRC (Information Technology Research Center) support program (IITP-2015-(H8501-15-1015)) supervised by the IITP (Institute for Information & communication Technology Promotion). The Corresponding author is Prof. Eui-Nam Huh.